\documentstyle[twocolumn]{jpsj}

\def\runtitle{Hydrodynamic Equation 
for the Breakdown of the Quantum Hall Effect in a Uniform Current}
\def\runauthor{Hiroshi {\sc Akera}}

\title{Hydrodynamic Equation for the Breakdown of the Quantum Hall Effect 
in a Uniform Current
\footnote{to be published in J. Phys. Soc. Jpn. {\bf 70} (2001) No.6.}}

\author{Hiroshi {\sc Akera}\footnote{E-mail address: akera@eng.hokudai.ac.jp}}

\inst{Department of Applied Physics, Hokkaido University, Sapporo 060-8628}

\recdate{October 16, 2000}

\abst
{
The hydrodynamic equation for the spatial and temporal 
evolutions of the electron temperature $T_e$ 
in the breakdown of the quantum Hall effect 
at even-integer filling factors in a uniform current density $j_x$ 
is derived from the Boltzmann-type equation, which takes into account  
electron-electron and electron-phonon scattering processes. 
The derived equation has a drift term, which is proportional to $j_x$ and 
the first spatial derivative of $T_e$.  
When applied to the spatial evolution of $T_e$ 
in a sample with an abrupt change of the width along the current direction, 
the equation gives a distinct dependence on the current direction 
as well as a critical relaxation,  
in agreement with results of recent experiments. 
}

\kword{integer quantum Hall effect, nonlinear transport, theory}

\begin{document}
\sloppy
\maketitle

\def\vecj{\mib j}
\def\vecE{\mib E}
\def\vecr{\mib r}
\def\vecq{\mib q}
\def\vecnabla{\mib \nabla}
\def\ve{\varepsilon}
\def\lvh{l_{\rm vh}}
\def\vt{\tilde v}
\def\eaia{\ve_{\alpha i a}}
\def\eaib{\ve_{\alpha i b}}
\def\ebkc{\ve_{\beta k c}}
\def\ebld{\ve_{\beta l d}}
\def\ebkd{\ve_{\beta k d}}
\def\faia{f_{\alpha i a}}
\def\faib{f_{\alpha i b}}
\def\fbkc{f_{\beta k c}}
\def\fbld{f_{\beta l d}}
\def\fbkd{f_{\beta k d}}
%

The diagonal conductivity $\sigma_{xx}$ is extremely small 
in the quantum Hall effect~\cite{Klitzing80,Kawaji81} (QHE) 
at small currents.
When the current is increased up to a critical value, 
$\sigma_{xx}$ increases by several orders of magnitude 
within a narrow range of current, 
and the QHE breaks down~\cite{Ebert83,Cage83,Kuchar84} 
(reviews are given in refs.\ \citen{Kawaji94,Nachtwei99,Komiyama00}).  
The hysteresis~\cite{Ebert83,Cage83,Kuchar84} 
with the change of the current 
as well as 
the variety of spatial\cite{Komiyama96,Kawaguchi97,Kaya98,Kaya99} 
and temporal\cite{Cage83,Ahlers93} behaviors 
observed in the breakdown of the QHE 
can be understood to be caused by 
a nonequilibrium phase transition. 
One of the most promising theories 
for describing the nonequilibrium phase transition 
is a hot-electron theory,~\cite{Ebert83,Gurevich84,Komiyama85}  
in which the change of $\sigma_{xx}$ is ascribed 
to the change of the electron temperature $T_e$.   
In this research, the spatial evolution of $\sigma_{xx}$ in the current direction 
is studied on the basis of the hot-electron theory.

The spatial evolutions of $\sigma_{xx}$ have been studied 
experimentally\cite{Komiyama96,Kawaguchi97,Kaya98,Kaya99} 
in a system with an abrupt change of the sample width at $x=0$ 
(the $x$ axis is taken along the current), 
which is designed to produce the change of 
the current density $\vecj=(j_x,0)$ at $x=0$. 
When the sample is designed so that  
$|j_x(x<0)| < j_{c1}<j_{c2}<|j_x(x>0)|$, 
with $j_{c1}$ and $j_{c2}$ the critical current densities 
(the system is bistable at $j_{c1}<|j_x|<j_{c2}$), 
the relaxation in $x>0$ of $\sigma_{xx}$ to the larger stable value  
and the development of the breakdown have been 
observed.\cite{Komiyama96,Kawaguchi97,Kaya99}
On the other hand, 
when $|j_x(x<0)| > j_{c2}>j_{c1}>|j_x(x>0)|$,  
the relaxation in $x>0$ of $\sigma_{xx}$ to the smaller stable value  
and the recovery of the QHE have been observed.\cite{Kaya98}
A unique feature in this nonequilibrium phase transition is that 
the length scale for the relaxation of $\sigma_{xx}$ 
depends strongly on the current direction: 
it is much larger for the electron flow in the $+x$ direction. 
It is only for this direction of the electron flow that  
the relaxation length in $x>0$
diverges as $j_x(x>0)$ approaches 
a critical value.\cite{Kawaguchi97,Kaya98,Kaya99}   
According to the hot-electron theory, 
this current-direction dependence 
can be understood by the drift motion of hot electrons along 
the current.\cite{Komiyama96,Kaya98} 
In order to understand the experiments 
and other spatial and temporal evolutions in a unified manner, 
the general theoretical framework is highly desirable. 

In this study, we derive the hydrodynamic equation 
describing the spatial and temporal evolution of $T_e$ 
from the Boltzmann-type equation 
on the basis of the microscopic scattering processes 
proposed previously.\cite{Akera00} 
We restrict our derivation to 
even-integer filling factors $2N$ ($N=1,2,\cdots$), for simplicity, 
and assume a uniform current distribution. 
We apply the obtained hydrodynamic equation to calculate 
the spatial relaxation of $T_e$ in the current direction.  
Our calculation reproduces both the current direction dependence 
and the critical behavior of the relaxation length. 

From a microscopic point of view, 
the potential in the plane of the two-dimensional electron system (2DES)
fluctuates due to the presence of the layer 
with ionized donors,~\cite{Nixon90} 
and the average distance between a potential hill and a neighboring 
valley, $\lvh$, is  
$\lvh \sim 0.1\ {\rm \mu m} \gg \ell$ with $\ell$ the magnetic length. 
The screened potential has an amplitude 
approximately equal to the Landau level separation,~\cite{Wulf88,Efros88} and
electrons (holes) populate the $N+1$th ($N$th) Landau level, 
even with a filling factor of $2N$. 
Electron orbits in such potential 
consist of localized closed orbits around a potential hill (valley) 
and extended orbits at the center of the broadened Landau level.
Wave packets constructed from extended states perform 
drift motions perpendicular to the electric field. 
In addition, 
an electron hops between a large hill orbit and a large valley orbit 
near an extended orbit. 
Such a hopping motion is given by an electron-electron scattering,  
in which one electron hops between a hill and a valley 
and, at the same time, 
another is excited or relaxed within a hill or a valley. 
Hopping motions along the electric field excite electrons, 
resulting in the energy gain of the 2DES. 
Electron-phonon scatterings, on the other hand, 
relax the 2DES, resulting in the energy loss.  
In our previous study,\cite{Akera00} 
the energy gain and loss have been calculated 
for the higher $T_e$ branch and the vanishing lattice temperature 
in homogeneous states.   
The theory is 
generalized in this study to include spatial variations of $T_e$  
in a macroscopic scale (much larger than $\lvh$) 
as well as an effect due to the drift motion of electrons in extended orbits 
(this effect is absent in homogeneous states).

In constructing the Boltzmann-type equation, 
the 2DES is divided into regions 
separated by an equipotential line of the fluctuating potential 
so that each region contains one hill or one valley. 
The distribution of electrons is described by 
the occupation probability $f_{\alpha i a}(t)$ 
of the $a$th state 
in the $i$th region at position $\vecr_i$ 
in level $\alpha$, 
where $\alpha$ is a combination of the Landau index and the spin and 
takes one of 
$N\uparrow, N\downarrow, N+1\uparrow$, or $N+1\downarrow$. 
The Boltzmann-type equation is then given to be 
\begin{equation}
{\partial \faia \over  \partial t}= 
\left({\partial \faia \over \partial t}\right)_{\!\!\rm drift}
+\left({\partial \faia \over \partial t}\right)_{\!\!\rm coll}
\ ,
\label{eq:Boltzmann}
\end{equation}
consisting of the drift term due to extended states and 
the collision term due to electron-phonon 
and electron-electron interactions. 
In electron-phonon scatterings, 
we have only included transitions $\alpha i a \rightarrow \alpha i b$
within the same region ($i$) within the same Landau level ($\alpha$), 
since their rates are much larger than others due to larger overlaps of 
wave functions.\cite{Akera00} 
In electron-electron scatterings, 
we have only included transitions 
$\alpha i a \rightarrow \alpha j b,\ \beta k c \rightarrow \beta l d$
within the same Landau level ($\alpha$, $\beta$). 
In the following we consider two classes of 
dominant electron-electron scatterings: 
(e1) $i=j$ and $k=l$, 
(e2) $i=j$, and $k$ and $l$ are the nearest neighbors.

The local equilibrium is assumed in the scale of $\lvh$, 
since the inelastic scattering length due to electron-electron scatterings 
is estimated to be $\sim \ell \ll \lvh$. 
Introducing the electron temperature $T_e(\vecr_i,t)$ and 
the chemical potential $\mu(\vecr_i,t)$ of the $i$th region,  
the occupation probability is given by
\begin{equation}
f_{\alpha i a}(t)= f(\ve_{\alpha i a};\mu(\vecr_i,t), T_e(\vecr_i,t)) ,
\end{equation}
where 
$
f(\ve;\mu, T_e) =1/\{\exp[(\ve-\mu)/k_B T_e] +1\}
$,
and $\ve_{\alpha i a}$ is the energy of state $\alpha i a$.

A state in the local equilibrium is 
specified by the electron temperature $T_e(\vecr,t)$, 
the chemical potential $\mu(\vecr,t)$, 
and the electric field $\vecE(\vecr,t)$ 
(these are assumed to be slowly varying 
in the scale of $\lvh$). 
The hydrodynamic equations for these variables are 
coupled in general. 
Here we assume that the current density is uniform, 
neglecting deviations near the sample edges. 
Although the current distribution in the Hall bar geometry 
is not uniform in the QHE,\cite{MacDonald83}
it is suggested to be uniform in the breakdown region by 
the observations in many samples~\cite{Kawaji93,Boisen94} 
that the critical current is proportional to the sample width. 
$\vecE$ is then given by $\vecE=(\rho_{xx}j_x,\rho_{yx}j_x)$, 
with a constant current density $\vecj=(j_x,0)$. 
Since $\rho_{xx}$ is smaller than $\rho_{yx}$ 
by, at least, one order of magnitude
and $\rho_{yx}$ is quantized even in the breakdown regime, 
$\vecE$ can be approximated by $\vecE = (0,-j_x h/2e^2N)$.
The chemical potential is then at the middle point between 
the $N$th and the $N \!+\! 1$th Landau levels 
and $\vecnabla \mu=e\vecE$. 
Owing to the assumed uniformity of the current, 
the equation for $T_e(x,t)$ is decoupled from 
those for $\vecE$ and $\mu$. 

To obtain the equation for $T_e(x,t)$, 
we introduce the energy density $U(x,t)$ defined by 
\begin{equation}
U(x,t)= {1 \over S} \sum_{\alpha, i,a} 
(\ve_{\alpha i a}-\mu_i) f_{\alpha i a}(t), 
\end{equation}
with $\mu_i=\mu(\vecr_i)$.
An average is taken over regions within area $S\ (\gg \lvh^2)$ 
in which the variation of $T_e$ is negligible. 
The time derivative of $U(x,t)$ is given 
in terms of the electronic specific heat $C_e=\partial U/\partial T_e$ as 
\begin{equation}
{\partial U \over \partial t}=C_e {\partial T_e \over \partial t },
\end{equation}
and is also given using eq.\ (\ref{eq:Boltzmann}) as 
\begin{equation}
{\partial U \over \partial t}= 
\left({\partial U \over \partial t}\right)_{\!\!\rm drift}
\!\! +\left({\partial U \over \partial t}\right)_{\!\!\rm coll}^{\!\!\rm p}
\!\! +\left({\partial U \over \partial t}\right)_{\!\!\rm coll}^{\!\!\rm e1}
\!\! +\left({\partial U \over \partial t}\right)_{\!\!\rm coll}
^{\!\!\rm e2}.  
\end{equation}
The three collision terms are the electron-phonon term, and
the electron-electron terms (e1) and (e2), respectively.  

The drift term is given by 
\begin{equation}
\!\!\! \left({\partial U \over \partial t}\right)_{\!\!\rm drift} \!\!\!\!
=\! - {\partial j_{Ux}^{\rm d} \over \partial x},
\end{equation}
with the energy flux $\vecj_U^{\rm d}=(j_{Ux}^{\rm d},0)$:
\begin{equation}
j_{Ux}^{\rm d}= 2 \hbar\omega_c f_{\rm ex} j_{nx}^{\rm d1} ,
\end{equation} 
where $j_{nx}^{\rm d1}=eE_y/h$ is the electron flow 
averaged within $S\ (\gg \lvh^2)$
at the filling factor one, 
$f_{\rm ex}=1/[\exp(\hbar\omega_c/2 k_B T_e) +1]$ 
is the occupation probability of extended states, 
and $\omega_c$ is the cyclotron frequency. 
The electron-phonon term is given by
\begin{eqnarray}
\left({\partial U \over \partial t}\right)_{\!\!\rm coll}^{\!\!\rm p}&=&
{1 \over S}\sum_{\alpha,i,a,b} (\eaib-\eaia)  
\faia (1-\faib) W^{\rm p}_{\alpha i a\rightarrow i b} \nonumber \\
&=&-P_L,
\end{eqnarray}
where $W^{\rm p}_{\alpha i a\rightarrow i b}$ 
is the transition rate and 
$P_L$ is the energy loss per unit area per unit time.
When $|\eaia-\eaib| \ll k_B T_e$ 
(this inequality holds approximately 
in the higher $T_e$ branch ($T_e \geq T_{c1}$)
since $|\eaia-\eaib| \sim 0.1 \hbar\omega_c$ and 
$k_B T_{c1}=0.324\hbar\omega_c$\cite{Akera00}), 
we have $\faia (1-\faib) \approx \faia (1-\faia) = -k_B T_e f'(\eaia)$   
and obtain 
\begin{equation}
P_L=C_L \tilde T_e,
\end{equation}
with $\tilde T_e=k_B T_e/ \hbar\omega_c$
and 
\begin{equation}
C_L={\hbar \omega_c \over S }
\sum_{\alpha,i,a,b} [- f'(\eaia)](\eaia-\eaib) 
W^{\rm p}_{\alpha i a\rightarrow i b} .
\end{equation}
The electron-electron term (e1) is given, 
when $\partial T_e / \partial x \ll T_e/\lvh$, by
\begin{equation}
\left({\partial U \over \partial t}\right)_{\!\!\rm coll}^{\!\!\rm e1}=
{\partial  \over \partial x} \kappa_0 {\partial T_e \over \partial x},
\end{equation}
where $\kappa_0$ is the thermal conductivity due to (e1) processes.  
When $|\eaia-\eaib| \ll k_B T_e$, $\kappa_0$ is given by 
\begin{equation}
\kappa_0 
\!=\! {k_B \over 8S} 
\!\!\!\!\!\! \sum_{\alpha,\beta,i,k,a,b,c,d} \!\!\!\!\!\!
W^{\alpha i a \rightarrow i b}_{\beta k c\rightarrow k d}
 [- f'(\eaia) ][- f'(\ebkc)] \Delta \ve^2  \Delta r^2 ,
\end{equation}
with $\Delta \ve=\eaia-\eaib$, $\Delta r=|\vecr_i-\vecr_k|$, and 
$W^{\alpha i a \rightarrow i b}_{\beta k c\rightarrow k d}$ 
the transition rate.
The electron-electron term (e2) is given by 
\begin{equation}
\left({\partial U \over \partial t}\right)_{\!\!\rm coll}^{\!\!\rm e2}
= P_G  - {\partial j_{Ux}^{\rm h} \over \partial x},
\end{equation}
where $P_G= (-e) \vecE \cdot \vecj_n^{\rm h}$ is 
the energy gain per unit area per unit time,   
and $\vecj_n^{\rm h}$ and $\vecj_U^{\rm h}=(j_{Ux}^{\rm h},0)$ are 
the electron flux and 
the energy flux due to hopping processes, respectively. 
When 
$|\eaia-\eaib| \ll k_B T_e$ and 
$eE \lvh \ll k_B T_e$
(the second inequality holds at $E \sim E_{c1}$
since $E_{c1}=20$V/cm ($B=5\ $T)\cite{Akera00}
and $eE_{c1}\lvh = 0.03 \hbar \omega_c$),  
we obtain 
\begin{equation}
P_G=C_G f_{\rm ex} (1-f_{\rm ex}) \tilde E^2,
\end{equation}
where $\tilde E=eE\ell / \hbar\omega_c$  ($\ell$ the magnetic length) 
and
\begin{equation}
C_G={(\hbar \omega_c)^2 \lvh^2  \over 4 \ell^2 S } 
\sum_{\alpha,\beta,i,k,l,a,b,c,d}
W^{\alpha i a \rightarrow i b}_{\beta k c\rightarrow l d} [- f'(\eaia)]  .
\end{equation}
Similarly we obtain 
\begin{equation}
j_{Ux}^{\rm h}
= - C_G  f_{\rm ex} (1-f_{\rm ex}) {\ell^2 \over 4 T_e} 
 {\partial T_e \over \partial x} .
\end{equation}

Finally we obtain the equation for $T_e$: 
\begin{equation}
C_e {\partial T_e \over \partial t }= 
{\partial \over \partial x}\kappa (T_e) {\partial T_e \over \partial x} 
- {\partial j_{Ux}^{\rm d} \over \partial x}
+P_G -P_L ,
\end{equation}
where the total thermal conductivity $\kappa$ is 
\begin{equation}
\kappa (T_e) =\kappa_0 
+ C_G  f_{\rm ex} (1-f_{\rm ex}) {\ell^2 \over 4 T_e} .
\end{equation}

When a constant density of states is assumed,  
$C_L$, $C_G$, and $\kappa_0$ are approximately independent of $T_e$. 
Estimates of $C_L$ and $C_G$ are given 
in our previous paper.\cite{Akera00} 
In estimating $C_L$, 
the deformation potential scattering by acoustic phonons and 
the vanishing lattice temperature were assumed;  
in estimating $C_G$, 
the summation over $k$ and $l$ was limited to the case 
where either $k$ or $l$ coincides with $i$. 
Calculated values of 
the critical electric field $E_{c1}\propto (C_L/C_G)^{1/2}$ 
and the energy dissipation $\sigma_{xx}E_{c1}^2 \propto C_L$ 
were in agreement with the observed values 
in their orders of magnitude. 
Following the calculation of $C_G$, 
here, we calculate $\kappa_0$,     
in which we limit the summation over $k$ to the nearest neighbors of $i$ 
since the contribution to $\kappa_0$ 
of larger $\Delta r$ decreases as $\Delta r^{-3}$. 
The estimation shows that 
$\kappa_0$ is only 5 percent of the total thermal conductivity, $\kappa$. 
The $T_e$ dependence of $\kappa$ is neglected in the following, 
since 
$\partial \kappa / \partial T_e=0$
at $E=E_{c1}$ and $T_e=T_{c1}$. 

Gurevich and Mints\cite{Gurevich84} 
introduced an equation for $T_e$ 
in their study of spatial and temporal variations of $T_e$ in the bistable region.
In their equation they included a term 
proportional to $j_x$ and $\partial T_e / \partial x$, 
which was ascribed to a contribution from the Seebeck effect. 
Although the drift term derived above is 
similar in form to their term 
[$\partial f_{\rm ex}/\partial x =(\partial f_{\rm ex}/\partial T_e)
(\partial T_e/\partial x) $], 
it has a different origin. 
The Seebeck effect vanishes 
in the present case of even-integer filling factors
because the density of states is symmetric with respect to the chemical 
potential.

In stationary states, the equation for $T_e$ becomes 
\begin{equation}
M (\tilde T_e)  {d^2 \tilde T_e \over d \tilde x^2}= 
\eta (\tilde T_e) {d \tilde T_e \over d \tilde x} 
+\tilde T_e -\gamma \tilde j_x^2 f_{\rm ex} (1-f_{\rm ex}) ,
\end{equation}
in a dimensionless unit 
where 
\begin{equation}
M (\tilde T_e)=\kappa \hbar\omega_c / k_B C_L \ell^2 ,
\end{equation}
\begin{equation}
\eta (\tilde T_e)=
\eta_0   \tilde j_x f_{\rm ex} (1-f_{\rm ex}) \tilde T_e^{-2},
\end{equation}
\begin{equation}
\eta_0= {eE_{c1} \omega_c / 2\pi \ell C_L } ,
\label{eq:eta0}
\end{equation}
$\tilde x=x/\ell$, 
$\tilde j_x=(-j_x)/j_{c1}=E_y/E_{c1}$, and $\gamma=2.23342$. 
The estimation gives $\eta_0=50$ at $B=5\ $T.
By changing the variable as $\tilde t=-\tilde x$, 
the above equation becomes the equation for 
``position" $\tilde T_e$ 
as a function of ``time" $\tilde t$ 
of a particle with mass $M (\tilde T_e)$ 
in a position-dependent friction $\eta (\tilde T_e)$ 
and a potential $V(\tilde T_e)$ defined by 
\begin{equation}
- \partial V/ \partial \tilde T_e =
\tilde T_e -\gamma \tilde j_x^2 f_{\rm ex} (1-f_{\rm ex}) ,
\label{eq:V}
\end{equation}
and plotted in Fig.\ 1(a). 
We have chosen this direction of ``time" because 
$\tilde T_e$ with increasing $\tilde x$ approaches a stable value 
in a simple exponential form and is convenient for the initial point.

The current direction dependence is immediately obtained 
from the fact that $\eta \propto \tilde j_x$.  
The motion of the fictitious particle is much slower 
for the positive friction ($\tilde j_x>0$) than for the negative friction 
($\tilde j_x<0$), 
reproducing the observed dependence of $T_e(x)$ 
on the current direction. 

A critical behavior is expected to occur at $|j_x|=j_{c1},\ j_{c2}$. 
We consider in this study the critical behavior at $|j_x|=j_{c1}$, 
since $j_{c2}=\infty$ in our simplified calculation of $P_G$ and $P_L$. 
At $|j_x|=j_{c1}$, the homogeneous steady solutions are 
$\tilde T_e=0$ and $\tilde T_e=\tilde T_{c1}=0.324$. 
We consider below  
a relaxation from $T_e> T_{c1}\ (x=0)$ to $T_e=0\ (x=\infty)$ 
when $|j_x| < j_{c1}$ ($|\tilde j_x|<1$).  

First we show the critical behavior qualitatively  
using the above equivalent mechanical system.   
The initial ``position" is chosen to be  
$\tilde T_{\rm init} \equiv \tilde T_e(\tilde x_{\rm init}) \ll \tilde 
T_{c1}$. 
At $\tilde T_e \sim \tilde T_{\rm init}$, $f_{\rm ex} \approx 0$ and 
$\tilde T_e = \tilde T_{\rm init} \exp[-(\tilde x -\tilde x_{\rm 
init})/ M^{1/2}]$. 
Therefore the initial condition at $\tilde x =\tilde x_{\rm init}$
is $\tilde T_e = \tilde T_{\rm init}$ and 
$d \tilde T_e / d \tilde t = \tilde T_{\rm init} / M^{1/2} >0$,
and the ``particle" goes down 
the slope of the potential as shown in Fig.\ 1(a). 
When friction $\eta$ is positive ($\tilde j_x>0$), 
the motion of the particle slows down 
at the gradual slope at $\tilde T_e \sim \tilde T_{c1}$ 
and it takes a longer time to pass through the slope  
as $\tilde j_x \rightarrow 1$
since $ \partial V/ \partial \tilde T_e \rightarrow 0$ ($\tilde j_x \rightarrow 1$). 
This means the divergence of the relaxation length of $T_e(x)$ 
as $\tilde j_x \rightarrow 1$. 
On the other hand, no such critical behavior occurs when $\tilde j_x<0$ 
since the friction is negative. 
\begin{figure}
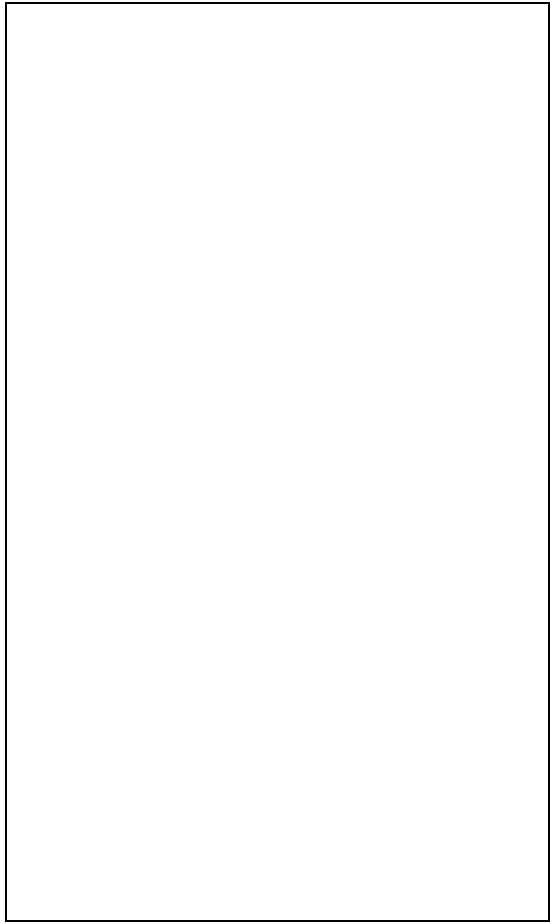

\figureheight{12cm}
\caption{
(a) ``Potential" $V(\tilde T_e)$ defined in eq.\ (\ref{eq:V}) 
($\tilde T_e=k_B T_e/ \hbar\omega_c$, $\tilde j_x=(-j_x)/j_{c1}=E_y/E_{c1}$). 
(b) $\tilde T_e$ as a function of $\tilde x =x/\ell$ 
with $\eta_0$ defined in eq.\ (\ref{eq:eta0}) 
(thicker lines: numerical solutions of eq.\ (\ref{eq:xT}), 
thinner lines: approximate analytical solutions, eq.\ (\ref{eq:analy})). 
(c) Relaxation length $L_s$ defined in eq.\ (\ref{eq:Ls}) 
as well as $h_0$, $h_2$, and $\tilde T_0$ 
defined in eq.\ (\ref{eq:T0h0h2}). 
} 
\end{figure}

Next we calculate the length scale of the slow variation of $T_e(x)$. 
In this calculation, 
the second-derivative term can be neglected  
because the decay length $l_d$ of 
$d^2 \tilde T_e/ d \tilde x^2$ is 
given by $l_d=\ell M/\eta$ 
when $\partial V/ \partial \tilde T_e = 0$  
and is estimated to be $l_d=5\ \mu$m
at $B=5\ $T and $T_e =T_{c1}$. 
The equation is integrated to give 
\begin{equation}
{\tilde x -\tilde x_0 \over \eta_0}=
- \tilde j_x  \int^{\tilde T_e}_{\tilde T_0} d \tilde T 
{1 \over h(\tilde T, \tilde j_x)} ,
\label{eq:xT}
\end{equation}
with 
$
h(\tilde T, \tilde j_x)= 
\tilde T^3 (e^{1/2\tilde T} + e^{-1/2\tilde T}+2) 
-\gamma \tilde j_x^2 \tilde T^2
$.
$\tilde T_0=\tilde T_e(\tilde x_0)$ is the temperature 
at which $h(\tilde T, \tilde j_x)$ takes its minimum. 
The numerical integration gives $\tilde T_e(\tilde x)$ 
plotted by thicker lines in Fig.\ 1(b),  
which shows that the length of the slow-variation region $L_s$ increases 
as $\tilde j_x \rightarrow 1$.
To find an analytical formula for $L_s$, 
we replace $h(\tilde T, \tilde j_x)$ by 
its Taylor series with respect to $\tilde T$ 
around its minimum up to the second order: 
\begin{equation}
h(\tilde T, \tilde j_x) \approx 
h_0(\tilde j_x) + h_2(\tilde j_x) [\tilde T -\tilde T_0(\tilde j_x) ]^2 ,
\label{eq:T0h0h2}
\end{equation}
with $h_0(\tilde j_x)$, $h_2(\tilde j_x)$, and $\tilde T_0(\tilde j_x) $ 
plotted in Fig.\ 1(c). 
The analytical integration gives 
\begin{equation}
\tilde T_e -\tilde T_0 = - \left({h_0 \over h_2}\right)^{1/2} 
\tan \left[{(h_0 h_2)^{1/2} \over \eta_0 \tilde j_x} (\tilde x -\tilde x_0) 
\right]  , 
\label{eq:analy}
\end{equation}
which is plotted in Fig.\ 1(b) by thinner lines; 
the curve is in good agreement with the numerical ones.  
The relaxation length $L_s$ is then defined by 
\begin{equation}
L_s /\ell =  \pi \eta_0 \tilde j_x (h_0 h_2)^{-1/2} .
\label{eq:Ls}
\end{equation}
$L_s$ diverges as $(1-\tilde j_x)^{-1/2}$ as shown in Fig.\ 1(c),  and
becomes $\sim 10\ \mu$m at $\tilde j_x=0.97$. 

The observed diagonal resistivity $\rho_{xx}$\cite{Kaya98}
remains to take large values 
within distance $L_s^{\rm exp}$ of approximately 50$\ \mu$m 
($\tilde j_x \approx 0.97$ and $B=4.8\ $T)
from the position of the hot-electron injection, 
and then undergoes a steep decrease. 
The calculated relaxation length, $L_s$, of $T_e$ 
(corresponding to $\rho_{xx}$)
is in agreement with the observed relaxation length, $L_s^{\rm exp}$, 
in the order of magnitude and the critical behavior. 
The present theory under the assumption of $T_e \sim T_{c1}$ 
is not applicable to 
the observed steep decrease 
in the lower-$\rho_{xx}$ (lower-$T_e$) region. 

In conclusion, 
we have derived the hydrodynamic equation 
for the electron temperature $T_e$
from the Boltzmann-type equation. 
We have applied the obtained equation to 
the spatial evolution of $T_e$ 
and have reproduced two observed results: 
the current direction dependence 
and the critical behavior of the relaxation length for $T_e$. 

The author would like to thank 
Y.\ Levinson, N.\ Tokuda, and Y.\ Asano for valuable discussions. 


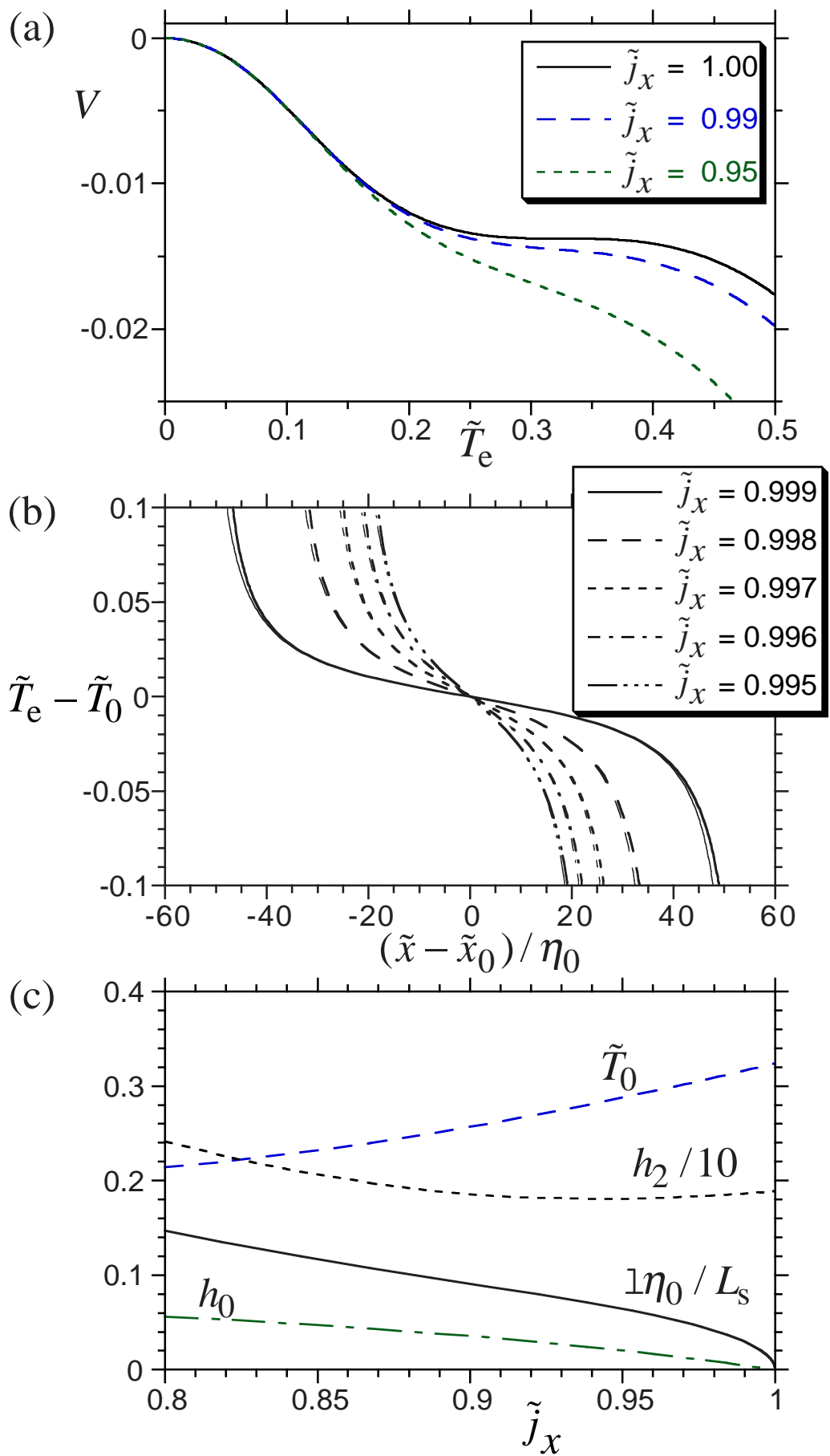
\end{document}